\documentclass[10pt]{article}
\usepackage{amsmath}
\usepackage{array}
\usepackage{placeins}
\usepackage{graphicx}
\usepackage[margin=1in]{geometry}
\usepackage[subtle]{savetrees}
\title{Surface Effects on the Magnetocrystalline Anisotropy of IrMn$_3$}
\author{Robert A. Lawrence, and Matt I.J. Probert}
\begin{document}
\maketitle

\begin{abstract}
   Magnetic anisotropy is a key parameter to describe the exchange bias effect in heterostructures. In this paper, we describe explicit density functional calculations of the magnetic structure of an interface between the industrially critically antiferromagnetic material, IrMn, and Fe, which together form a simple ferromagnet-antiferromagnet heterostructure. Additionally, the magnetic anisotropy was evaluated for several terminations of the IrMn. It was found that the [111] surface had a perpendicular anisotropy of 1.62 meV/$\text{\AA}^2$, whereas the two possible [100] surfaces (Ir-rich and Mn-rich) had in-plane anisotropies of 0.13 meV/$\text{\AA}^2$ and 1.39 meV/$\text{\AA}^2$ respectively. The affect of the magnetic order of the easy and hard configurations were calculated and used to explain the relative values of the anisotropies.
    
\end{abstract}

\section{Introduction}

A key property for spintronics applications is magnetocrystalline anisotropy \cite{Misiorny2013}. This intrinsic property, along with domain volume, controls the energy barrier for spin flipping of a domain. Soft magnetic materials -- those with a low barrier to reversal -- are key for applications such as low-loss power transformers where hard magnetic materials -- with a large barrier -- are critical for applications such as data storage, as well as more sophisticated spintronic devices such as magnetic tunnel junctions \cite{Samanta2024}.  For these junctions, biasing the orientation of a ferromagnetic layer using a hard antiferromagnet is key to their function, using a phenomenon known as the exchange bias effect\cite {Giri2011}.

The most important antiferromagnet in terms of usage is currently IrMn\cite{Frost2024}. Whilst widely used thanks to its high-performance, the cost of Ir is prohibitive to wide-scale use; the inelasticity of supply\cite{Vesborg2012} will lead to a drastic increase in cost given an increase in demand. Accordingly, gaining deeper insight into how an archetypal system such as IrMn works will enable more effective searching for Ir-free replacement materials with similar performance. The importance of IrMn for spintronic device manufacture \cite{Kang2021} has led to it being highly studied\cite{Carter2024}, yet there is a scarcity of theoretical work considering the most important part of a device; the interface between layers of different materials within a heterostructure. In this paper, we will explore the effect of interfaces on the magnetic structure of IrMn thin film layers, and consider how this affects the exchange bias \cite{Bufaial2024} that the IrMn is capable of inducing in a neighbouring soft ferromagnet\cite{Aley2010,Peng2020,Fan2022}.

A key controlling factor for maximising exchange bias device performance is the choice of interface between layers. Even for a fixed choice of materials which form the heterostructure, there are significant degrees of freedom in the choice of cleavage planes and this choice can have significant effects on the final performance of the heterostructure. This is due to the different combinations of cleavage planes leading to different surface states with noticeably different properties \cite{Aley2010}, including the magnetic ordering at the surface and the effective anisotropy that leads to the exchange bias. 

In order to draw greater insight into how the interfaces affect the magnetic properties, it is also important to understand the magnetic ordering in the limits of the bulk-phase and of a simple surface interacting with vacuum. These two cases are the extreme limits of the interface within a heterostructure: the bulk for the limit where there is no significant change, and the vacuum for the limit of maximal change. Accordingly, in addition to studying a simple heterostructure interface, we also consider the magnetic ordering for the surface case and bulk.

Determining magnetic ordering in an antiferromagnet at an interface is experimentally challenging; these layers are too thin to be resolved effectively using neutron diffraction or x-ray based techniques. Accordingly, computational insight from first principles simulations are a key technique to gain greater understanding of the magnetic ordering at the interfaces. In this work, density functional theory (DFT) will be used to perform electronic structure calculations, and thereby calculate the magnetic order without relying on empirically-derived parameters for the bulk material.

One major challenge when considering magnetic thin-films using DFT is that the spin-structure makes the self-consistent field (SCF) procedure into a multi-minimum problem and it becomes highly possible to converge to a stable ``ground state'' solution which can be in the order of 10 eV higher in energy than the true ground state magnetic order. This is conventionally resolved by a process of ``spin initialisation'', whereby extra information on the magnetic ordering -- typically derived from experiment -- is used to guide the SCF procedure into the correct minimum. This is not possible when experimental data does not exist, such as in the ultra-thin film limit, and accordingly we have developed a new method based upon a principle of minimum symmetry breaking to perform our initialisations.

To resolve this, we have developed a symmetry-based ansatz for the prediction of good initialisations that will lead to the correct magnetic ground state. This enables the prediction of magnetic structures from first principles without requiring any prior knowledge beyond the crystal structure of the system (which is a standard requirement for first-principles methods such as density functional theory).  

To begin this paper, we will discuss the ansatz for spin-initialisation created to perform this work (section \ref{sec:Epikernal}), and then proceed to consider the magnetic ordering in ultra-thin films (section \ref{sec:Bulk} onwards) and the impact of an interface with vacuum compared with a ferromagnetic capping layer with the same crystal structure but different elements. Finally, the magnetocrystalline anisotropy energy (MAE) of the system under rotation of the magnetic orientation of the ferromagnetic capping layer will be considered. To this end, we chose Fe, a very soft ferromagnetic material, in order to minimise any internal effects of rotating the magnetisation of the capping layer. This combination has also previously been of experimental interest \cite{Fan2022}. 

\section{Spin Structure Initialisation via the Epikernel Principle}\label{sec:Epikenal}
One of the major challenges of modelling thin films of IrMn$_3$ is the magnetic ordering. In principle, this is an emergent property of the system, however in practice the limitations of density functional approximations usually means that while the correct magnetic configuration will be the global minimum of energy, density functional calculations may readily converge to a local minimum instead.

To counteract this, standard practice is to initialise the spins of the unit cell -- thereby steering the SCF convergence towards the desired local minimum, usually the ground state magnetic structure. This procedure is simple and supported by many DFT codes, but requires \emph{a priori} knowledge of the magnetic structure of a system. This is usually not the case for the thin film limit (where the lack of sensitivity of neutron scattering can significantly limit our ability to resolve the magnetic structure) and especially not so for novel structures generated via high throughput screening as is becoming commonplace in materials discovery \cite{Bhattarai2025}.

In this section we will investigate the use of the ``Epikernel Principle'' \cite{Ceulemans1984} and how it may be used to predict magnetic ordering.

\subsection{The Epikernel Principle} \label{sec:Epikernal}

As originally stated by Ceulemans \emph{et al.}\cite{Ceulemans1984}, the epikernel principle states that the preferred symmetry-breaking direction for a Jahn-Teller unstable crystal is in the direction of a maximal epikernel \cite{Ascher1977}. We note, however, that the group theoretic approach Ceulemans \emph{et al.} use relies only on an applied operation (in their case, the atomic displacement that lowers the symmetry) and that the introduction of a magnetic moment to a site is also a symmetry-lowering operation. We also note that they use the term ``Jahn-Teller unstable'' to indicate partially unoccupied degenerate levels. These partially unoccupied degenerate levels may be energetically unstable with respect to spin polarisation as well as spatial polarisation (displacement). 

In the absence of spin-orbit coupling (SOC), magnetic polarisation does not affect the spatial orbitals -- and nor can the underlying symmetry of the crystal affect the magnetisation orientation. In the presence of SOC, however, this is not the case; $L$ and $S$ are no longer good quantum numbers and this magnetic polarisation is coupled to the lattice degrees of freedom giving rise to a magnetic anisotropy, known as the magnetocrystalline anisotropy. Now as well as a spin-magnitude (as can readily be predicted from Hund's rules), an orientation is also required to correctly model the magnetic system. 

For localised magnetic moments (i.e. not the itinerant magnetism limit), we note that the associated orbitals must also be tightly confined in space and approximately non-dispersive. In the purely localised moment limit, the orbitals themselves are also purely localised and therefore the symmetry of the electric potential that they are in may be well-determined purely by  knowledge of the local point group of that individual atom. We also know the symmetries of the degenerate orbitals under that point group in terms of their representation. This will be a reducible representation and, as with the Jahn-Teller case, we may apply the epikernel principle to predict the symmetry of the magnetic ground state.  

Now, for an \emph{arbitrary} set of degenerate levels, we note the epikernel principle -- that the system will be most stabilised by a symmetry breaking distortion that leaves a maximal epikernel of the original point group as the new point group of the reduced symmetry system. (An alternative expression is that the extremal points of the PES will exist at maximum epikernels, such that both the maximum energy and minimum energy symmetry breaking will occur at separate maximal epikernels). If a maximum energy maximal epikernel is chosen, noise within the SCF process will drive the system away from this maximum, which typically exhibits as single-site spin quenching, at which point a separate maximal epikernel may be chosen until a stable one is discovered. 

We note that by convention when dealing with point groups the local $\hat{z}$ direction is defined to be along the highest order rotation axis (e.g. a 6-fold rotation axis is  higher order than a 4-fold one). This is entirely independent of the global $\hat{z}$ direction, which is conventionally chosen to align with the $\vec{c}$ axis of the unit cell -- although they may be degenerate directions.

The final important concept to introduce is that spin is not a \emph{vector} but rather a \emph{pseudovector}. An arbitrary pseudovector $\vec{s}$  transforms under a transformation matrix $R$, such that $\vec{s}' = \text{det}(R)R\vec{s}$ -- and accordingly behaves the same under both proper and improper rotations (and is not affected by mirror planes). This leaves us with the result that we wish to choose the highest order rotation axis \emph{whether proper or improper} as our local $\hat{z}$ direction. In the case of IrMn$_3$, this is the 4-fold rotation axis normal to the face containing the Mn atom in question. 

In principle, the point group of an atom in an infinite crystal depends on all of the atoms in said crystal. In practice, however, the principle of near-sightedness \cite{Prodan2005} states that beyond a certain distance, no change will affect our atom. However, the work by Kohn \emph{et al.} does not provide a ready method to choose said distance. Recalling that we require only an \emph{approximation} to the correct spin-density that is sufficiently good as a spin initialisation, we note that the effects of the nearest neighbours will dominate the potential, and therefore elect to only consider nearest neighbour effects when defining the point group of a given atom. Whilst not strictly accurate, this has proved sufficient to converge from our initialisation towards the correct spin structure, as we shall now discuss.

\section{Methods}
Density functional theory was used within the plane-wave code CASTEP \cite{Clark2005} to simulate the magnetic structures and total energies of the systems used in this work. In all cases, the Perdew-Wang formulation \cite{Perdew1992,Perdew2018} of the LDA exchange-correlation functional was used.

\subsection{Bulk IrMn$_3$}

 In order to test the spin initialisation ansatz, bulk IrMn$_3$ was simulated, with both a conventional ``correct'' and ``ansatz-predicted'' spin initialisation. These were converged using a 2200 eV cut-off energy with the CASTEP NCP-19 library of norm-conserving vector-spin pseudopotentials, and a 20 $\times$ 20 $\times$ 20 Monkhorst-Pack grid was used for reciprocal space sampling of the 4-atom primitive cell. These both converged to the triangular frustrated structure in 20 SCF steps using a Pulay mixing scheme \cite{Pulay1980}. Other initialisations were attempted, and were found to converge to alternate, higher-energy (metastable) magnetic configurations or not converge. We speculate the successful converge of our ansatz came about because, whilst technically wrong, the ansatz yields a magnetic structure in the global minimum potential well within the wider global potential energy landscape, whereas other initialisations are closer to local minima, and thereby converge to higher-energy structures under a local minimisation. 

\subsection{Ultra-thin film IrMn$_3$}

For determining the spin ordering only, for the ultra-thin thin films a 12$\times$  12 $\times$ 1 Monkhorst-Pack grid was used with a 1200 eV cut-off-energy. The in-plane lattice parameters were 5.33$\text{\AA}$ for the [11] system and 3.77 $\text{\AA}$ for the [100]-terminated system. This was sufficient to converge the qualitative ordering under investigation. For the MAE calculations, further detailed in the next subsection, a 2200 eV cut-off energy and a 20 $\times$ 20$\times$ 1 k-point grid was used. A vacuum spacing of 10 $\text{\AA}$ was used throughout, with the spacing being increased on addition of the Fe layer. 

These systems then had a bilayer of Fe added on top, such that the underlying crystal structure of the IrMn$_3$ was preserved, in order to mimic the experimental set-up for determining anisotropy in antiferromagnets \cite{OGrady2020}. This system was not structurally relaxed to ensure that the changes to magnetic ordering were purely down to the change in the interface --from vacuum to Fe. Finally, the IrMn layers were removed and the difference in MAE for the bulk Fe recorded, which was several orders of magnitude lower in energy ($\mathcal{O}(\mu$eV)). 

\subsection{Exchange Bias Calculations}

Simulating the exchange bias effect directly has the twin advantages of being well-defined (one can deal with the magnetisation of the FM layer as a single order parameter rather than the multiple N\'eel vectors of a frustrated antiferromagnet)  and being more directly comparable with experimental results since the coupling across the interface is directly included. To simulate this, the thin layer of Fe with matching symmetry to the IrMn$_3$ layer has its magnetisation vector rotated through 90 degrees, and the energy difference of the entire system was recorded. The same rotation was performed with the Fe situated on top of the IrMn$_3$ ultra-thin films, and the change in the magnetocrystalline anisotropy energy (MAE) due to the IrMn$_3$ was then calculable -- this is the exchange bias of the system.

In order to ensure that the desired magnetic configuration for the Fe was sampled, the outermost (furthest from the interface) layer of Fe had its spin orientation constrained \cite{Cuadrado2018}, with the rest of the spins in the system free to relax to their lowest energy configuration under that constraint. This ensured sufficient freedom at the interface to ensure that the interfacial interactions were not artificially constrained.

\section{Bulk IrMn$_3$}\label{sec:Bulk}
\FloatBarrier
In order to test the ansatz on a crystal with a known, frustrated antiferromagnetic structure,  IrMn$_3$ -- which in the bulk phase is known to have a triangular magnetic ordering -- was used as a test case.
IrMn$_3$ is a cubic alloy of the Cu$_3$Au (L1$_2$) crystal structure, and belongs to space group 221 (Pm$\Bar{3}$m). 

\begin{figure}
    \centering
    \includegraphics[width=0.5\linewidth]{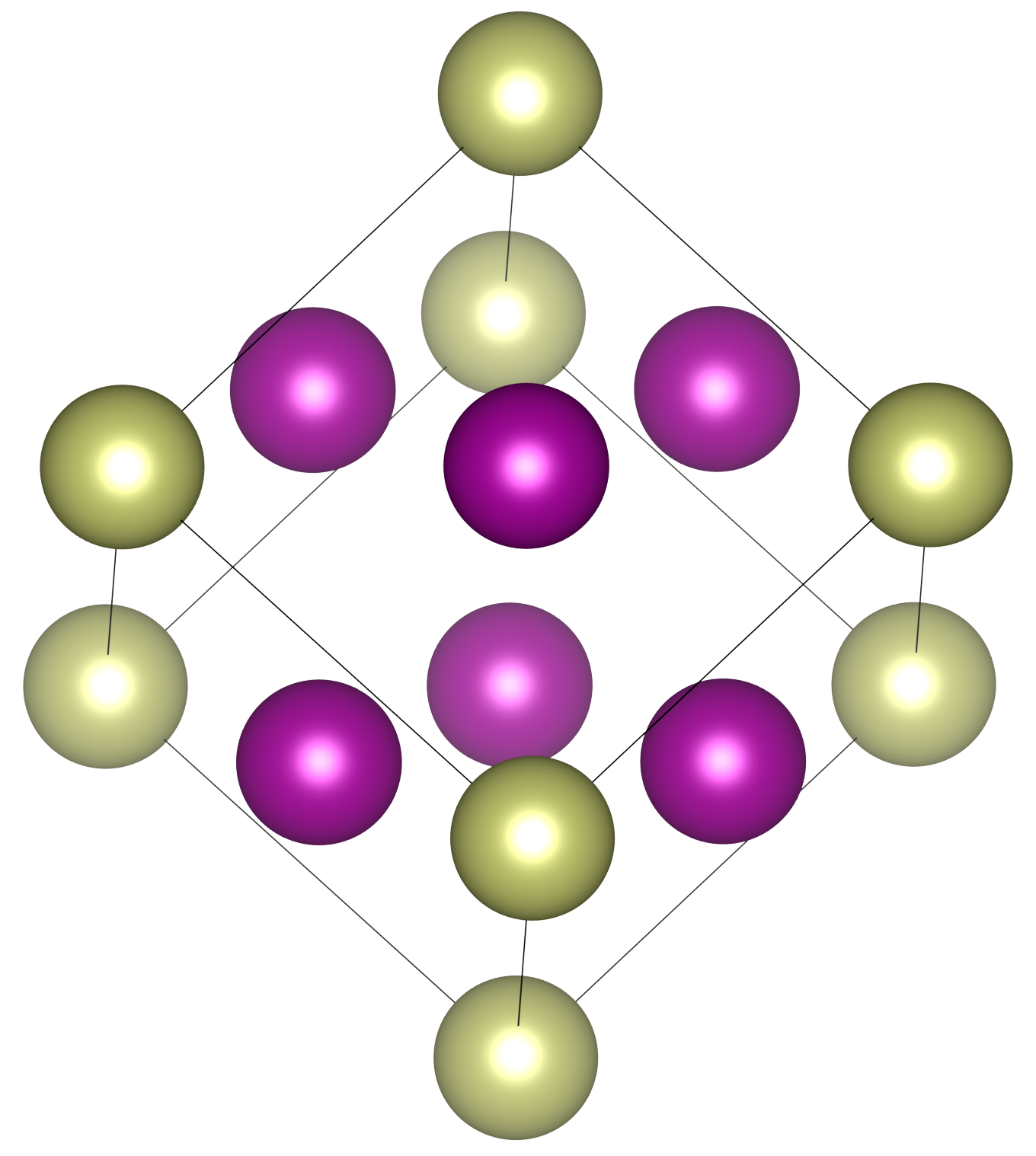}
    \caption{Bulk structure of IrMn$_3$. The [111] axis is aligned vertically in the page. Gold atoms are Ir and purple atoms are Mn.}
    \label{fig:enter-label}
\end{figure}

\begin{figure}[h!]
    \centering
    \includegraphics[width=0.4\linewidth]{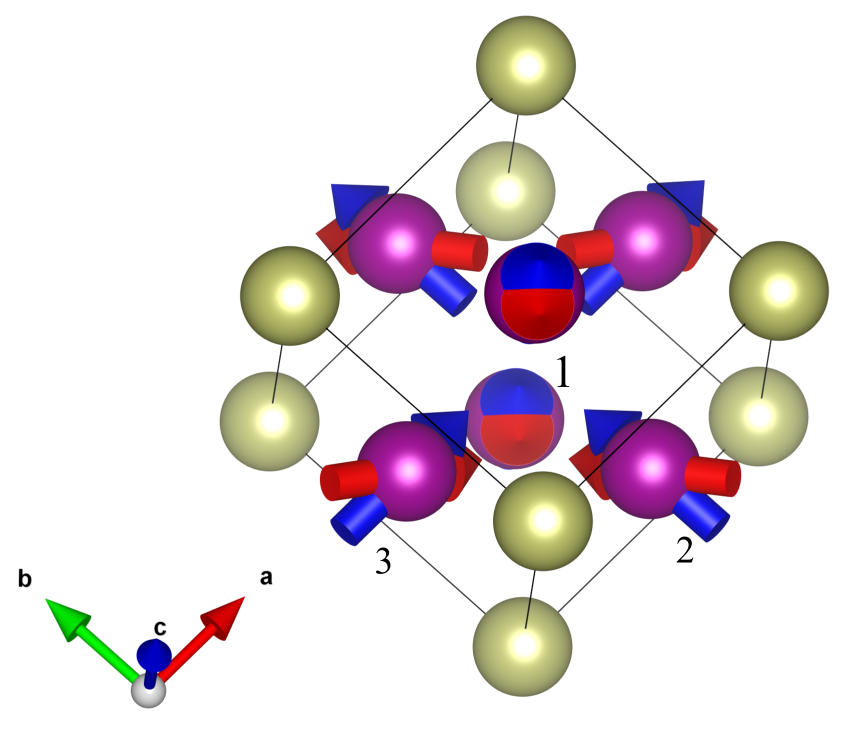}
    \caption{Bulk IrMn$_3$ including magnetic ordering. Purple atoms are Mn and bronze atoms are Ir Blue Arrows represent the ansatz's prediction, and red arrows represent the true spin ground state. Numbers next to the atoms are the label for each atom in table \ref{tab:spin_inits}. Note that the ansatz here provides approximately the same structure, saving only a net ferromagnetic moment along the $<111>$ direction.}
    \label{fig:BulkIrMn}
\end{figure}

This ansatz predicts the directions for the moments should be aligned along the normal of the face in which the Mn atom resides (see figure \ref{fig:BulkIrMn}). This is in agreement with the reported local easy axis directions by Szunyogh \emph{et al}. \cite{Szunyogh2009}. The directions of the initial orientation predicted by the ansatz and the true ground state configuration are both reported in fractional co-ordinates in table \ref{tab:spin_inits}.

\begin{table}[]
    \centering
    \begin{tabular}{c  c*{3}{>{\raggedleft\arraybackslash}p{3em}}  c*{3}{>{\raggedleft\arraybackslash}p{3em}} }
    \hline
     Mn Site   & \multicolumn{3}{r}{Ansatz Initialisation} & \multicolumn{3}{r} {Ground State Structure}  \\
     \hline
       1  & 0 & 0 & +1 &   $-\frac{1}{3}$  & $ -\frac{1}{3}$  & $+\frac{2}{3}$ \\ 
       2  & 0 & +1 & 0 &   $-\frac{1}{3}$  & $ +\frac{2}{3}$  & $-\frac{1}{3}$ \\ 
       3  & +1 & 0 & 0 &   $ +\frac{2}{3}$ & $ -\frac{1}{3}$  & $-\frac{1}{3}$ \\ 
       \hline
    \end{tabular}
    \caption{Ansatz-predicted and final state spin structures in fractional co-ordinates (normalised by spin magnitude). The known global ground state structure is conventionally used to initialise the magnetic moments in this cell. In practice, spin magnitudes are initialised with a magnitude of 5$\mu_B$, representing the use of Hund's rules to predict the spin.}
    \label{tab:spin_inits}
\end{table}

Two separate calculations were run, one with the correct ground state spin structure used as its spin initialisation, and another with the spin structure predicted using the ansatz. All other parameters, including the use of the Pulay mixing scheme, and the energy convergence tolerance of $1\times10^{-5}$ eV/atom were kept the same. These calculations found that the same number (20) of self-consistent field (SCF) cycles were required to attain the ground state for both initialisations, and these final structures were the same within 10 m$\mu_B$ magnetisation in any one direction. 

Finally, we note that this ansatz of aligning the local magnetisation with the highest order rotation axis of the local point group within a nearest-neighbour approximation \emph{for spin initialisation only} seems to be able to reproduce the experimentally confirmed global ground state magnetic structure with a high degree of accuracy for relatively complex frustrated systems. It does not, however, predict global spin-structures (such as FM or AFM structures) which is beyond the capability of a purely localised model.

For the thin-films we will investigate in the second part of this work, such an ansatz is vital -- determining the spin structure experimentally is difficult, even assuming that such thin films could be created, and therefore initialising to experimental values is not possible, and an exhaustive search of the spin-space, even coarsely, is not achievable due to the exponential scaling of the search space with number of atoms (for example, a 15 degree search mesh in $\theta$-$\phi$ space would require $288^{N_\text{atoms}}$ independent calculations with tightly converged numerical parameters to achieve).  

\FloatBarrier
\section{IrMn$_3$-Vacuum Interface}
\FloatBarrier

Firstly, we consider the effects of an interface of our crystal with vacuum on the underlying ground state spin structure of the crystal. 

\subsection{[100] Interface}

\begin{figure}
    \centering
    \includegraphics[width=0.7\linewidth]{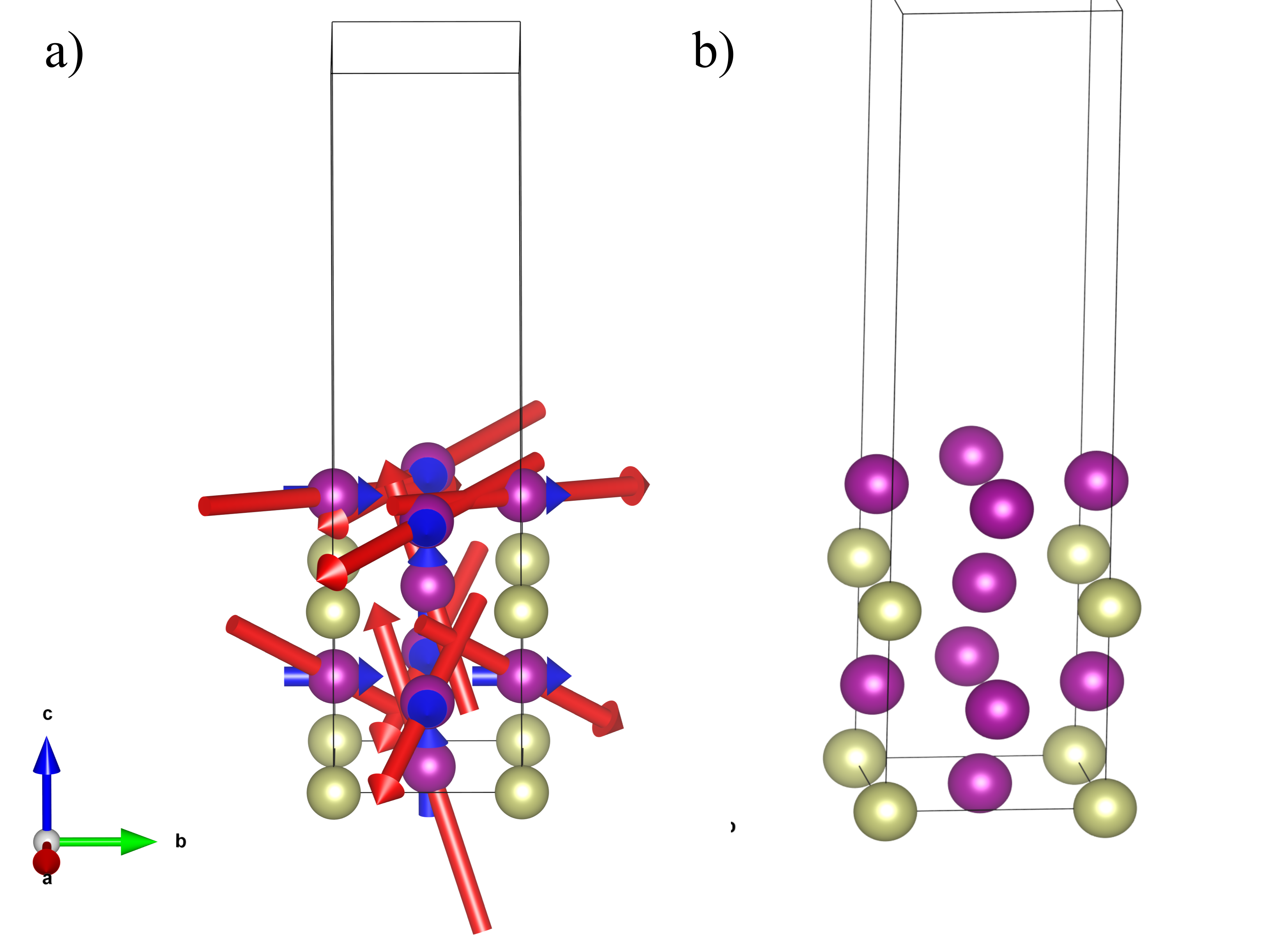}
    \caption{Panel a) Spin Structure of [100] IrMn$_3$ with vacuum. Panel b) Structure of [100] IrMn$_3$ with spins removed for clarity. Purple atoms are Mn and bronze atoms are Ir. Red arrows represent the spin structure, and blue arrows represent the spin initialisation. Note the different magnetic ordering between the two surface layers; the bottom (Ir and Mn capped) layer is highly resemblant of the bulk magnetic ordering, whereas the top (Only Mn capped) layer is highly distorted with a significant in-plane component. This is likely driven by the difference in nearest-neighbour defined environments.}
    \label{fig:100_vac}
\end{figure}

The first thing that may be seen by inspecting the magnetic structure shown in figure \ref{fig:100_vac} is that the magnetic ordering depends on which surface termination is present. Along the [100] axis, two possible terminations, either Mn-rich or Ir-rich are possible. At the Ir-rich termination, the nearest neighbour interactions for the Mn (which are with the Ir) are indistinguishable from bulk, leading to an approximate preservation of the triangular structure (with some perturbation due to the interfacial effects. At the Mn-rich interface, however, half of the nearest-neighbour atoms have been removed, with the effect that there is a significant rotation of the spins in the Mn-rich surface to being approximately coplanar with the interfacial layer.  

With all but the most careful epitaxial growth techniques, a real sample would likely have some combination of the different interfaces (for example if it is a terraced surface), and thus a device constructed using these interfaces would have properties that are some weighted average of the two separate surfaces. This average could be influenced by growth conditions (a Mn-rich growth would make Mn-rich termination more likely), and potentially by the effects of any capping layer, which will be further investigated in section \ref{sec:Fe-cap}.

\subsection{[111] Interface}

A more widely-used interface is the  [111] termination. Unlike the [100] case, the constructable terminations are identical, since all [111] planes vary only by a translation. Accordingly, there is only one possible interface that may be obtained from cleaving a bulk crystal, leading to an increased reliability for device purposes. This is also reflected in the magnetic structure, where all layers are (barring a translation) nearly the same (the external layers feature slightly larger magnetic moments). It is also worth noting that the [111] cell has a single moment that is shared between two of the triangular motifs, and unlike the bulk case, this atom has a notably longer magnetic moment. This would likely be associated with a stronger coupling to any capping layer's magnetic moments if it is robust against the introduction of the capping layer. 

\begin{figure}
    \centering
    \includegraphics[width=0.7\linewidth]{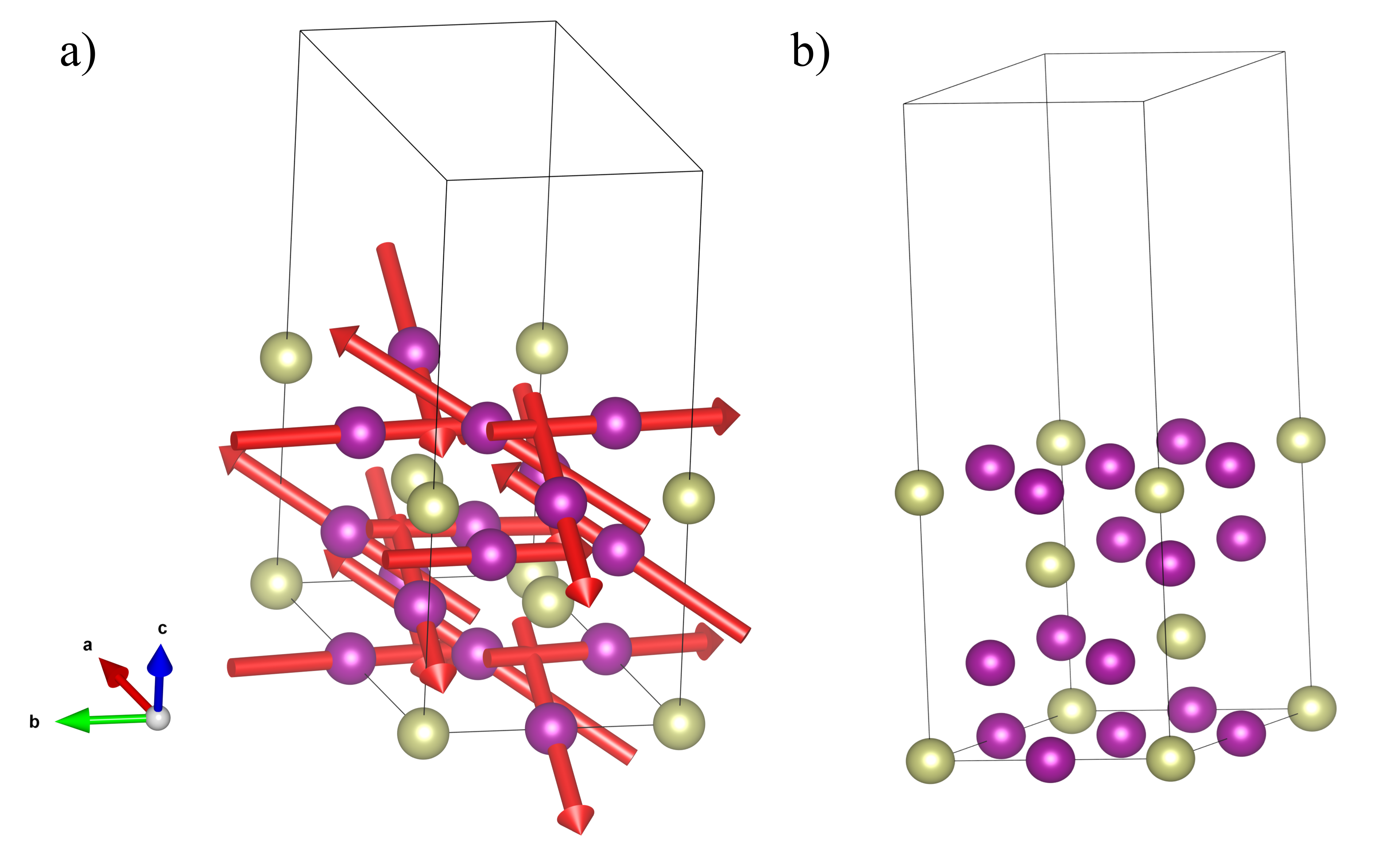}
    \caption{Panel a) Spin Structure of [111] IrMn$_3$ with vacuum. Panel b) Structure of [111] IrMn$_3$ surface with spins removed for clarity. Purple atoms are Mn and bronze atoms are Ir. Red arrows represent the spin structure, and blue arrows represent the spin initialisation. Note that the ``central'' Mn ions that are part of two triangles have a longer magnetic moment.}
    \label{fig:111_vac}
\end{figure}

\FloatBarrier
\subsection{Summary of Interfaces with Vacuum}

For both orientations, within the core of the layer the bulk spin structure is approximately conserved -- with the triangular motif present in this region. At the surfaces, however, the spin  structure is highly dependent on the symmetry of the termination. This is indicative of the localised moments being dominated by the nearest-neighbour environment. For the interior layers, the nearest-neighbour defined chemical environment is indistinguishable from bulk, and accordingly the final spin structure \emph{ought} to be very similar (but not identical, as there \emph{are} differences due to the thin-film termination). 

One may consider that the vacuum is in effect a region of high potential with respect to the interior of the crystal, with the difference being the work function. This means that there are large electric fields near the surfaces of the material compared to the interior of the crystal. The IrMn$_x$ family are metallic alloys, accordingly, they have a strong ability to screen these effects, thereby localising the changes to the surfaces of the crystal. For insulating systems, it is not clear that the changes to the magnetic structure should be so tightly confined to the surface layer, and further investigations should occur into the effects of surfaces of non-metallic systems. Nevertheless, it seems that this observation of vacuum interfaces only strongly affecting the magnetic ordering near the surface may be applicable to all metallic systems. 

As has been previously noted \cite{Jenkins2021}, there has been disagreement within the literature over the spin structure of IrMn$_3$. Some experimental references refer to the spin structure as being collinear \cite{OGrady2010}, whereas theoretical modelling \cite{Szunyogh2009} has determined that bulk IrMn$_3$ has the triangular spin structure reported in section \ref{sec:Bulk}. One feature of our results for the spin-structures of thin-film IrMn$_3$ is that the surface-layer has either approximately collinear ordering ($<100>$-termination) or one spin that has a significantly longer moment than the others ($<111>$-termination), both of which could potentially be mistaken for a long-range collinear-ordering when being analysed with experimental methods. This also confirms the stability of the triangular magnetic ordering for bulk IrMn even in the ultra-thin film limit.

\FloatBarrier
\section{IrM$_3$-Fe Interface}\label{sec:Fe-cap}

Devices for practical applications typically do not rely on an interface with vacuum. Accordingly, in the following section we consider the effects of capping our previous layers with a thin film of Fe, both to investigate the robustness of the ground-state magnetic structures we previously determined against capping and to evaluate the effect of the termination orientation on induced anisotropy in a neighbouring magnetically soft material.

\subsection{[100] Interface}
\FloatBarrier

\begin{figure}
    \centering
    \includegraphics[width=0.5\linewidth]{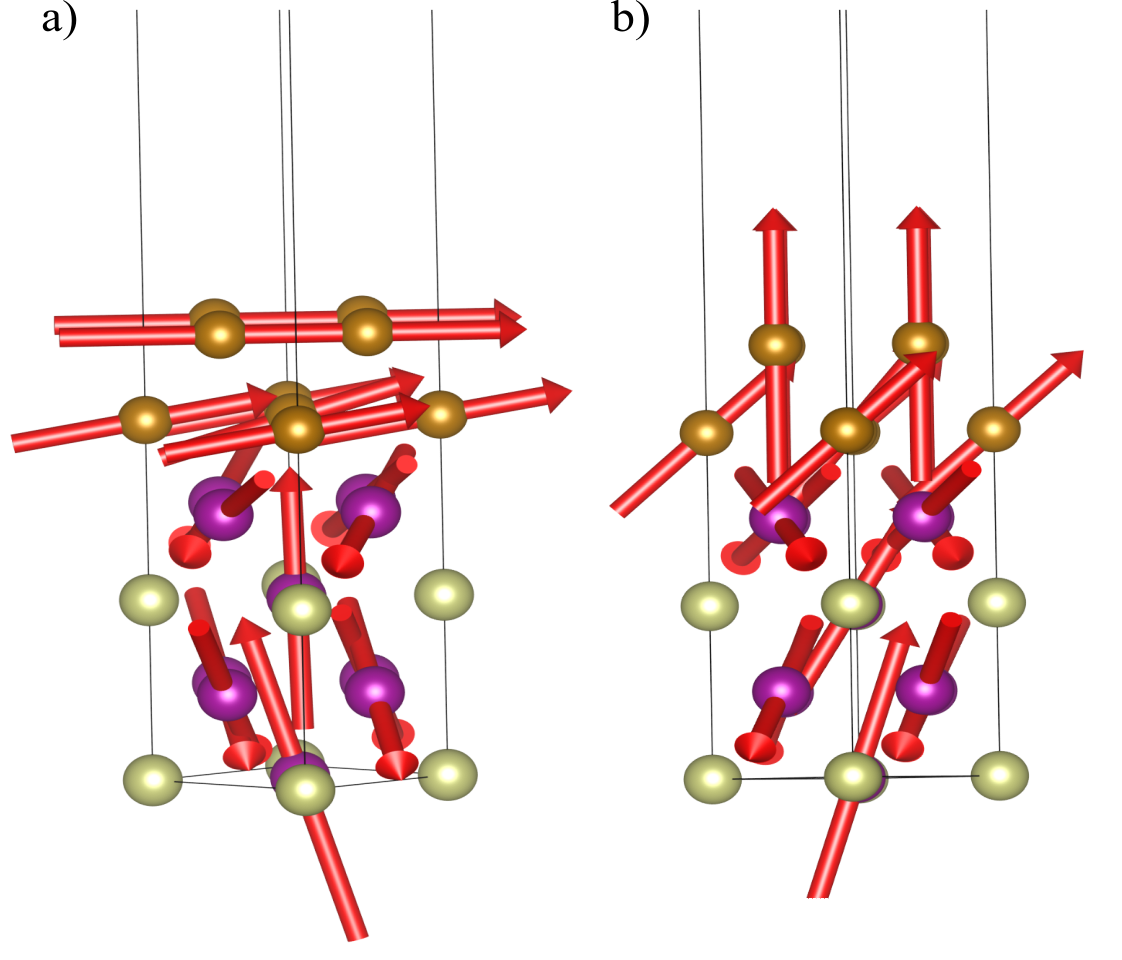}
    \caption{Magnetic structures for the a) Easy and b) Hard magnetic configurations for Fe on the Mn-rich [100] IrMn surface. Note the easy configuration most closely preserves the magnetic ordering in both the ferromagnetic Fe layer and the antiferromagnetic IrMn layer.}
    \label{fig:100_Fe_spins}
\end{figure}

\begin{figure}
    \centering
    \includegraphics[width=0.5\linewidth]{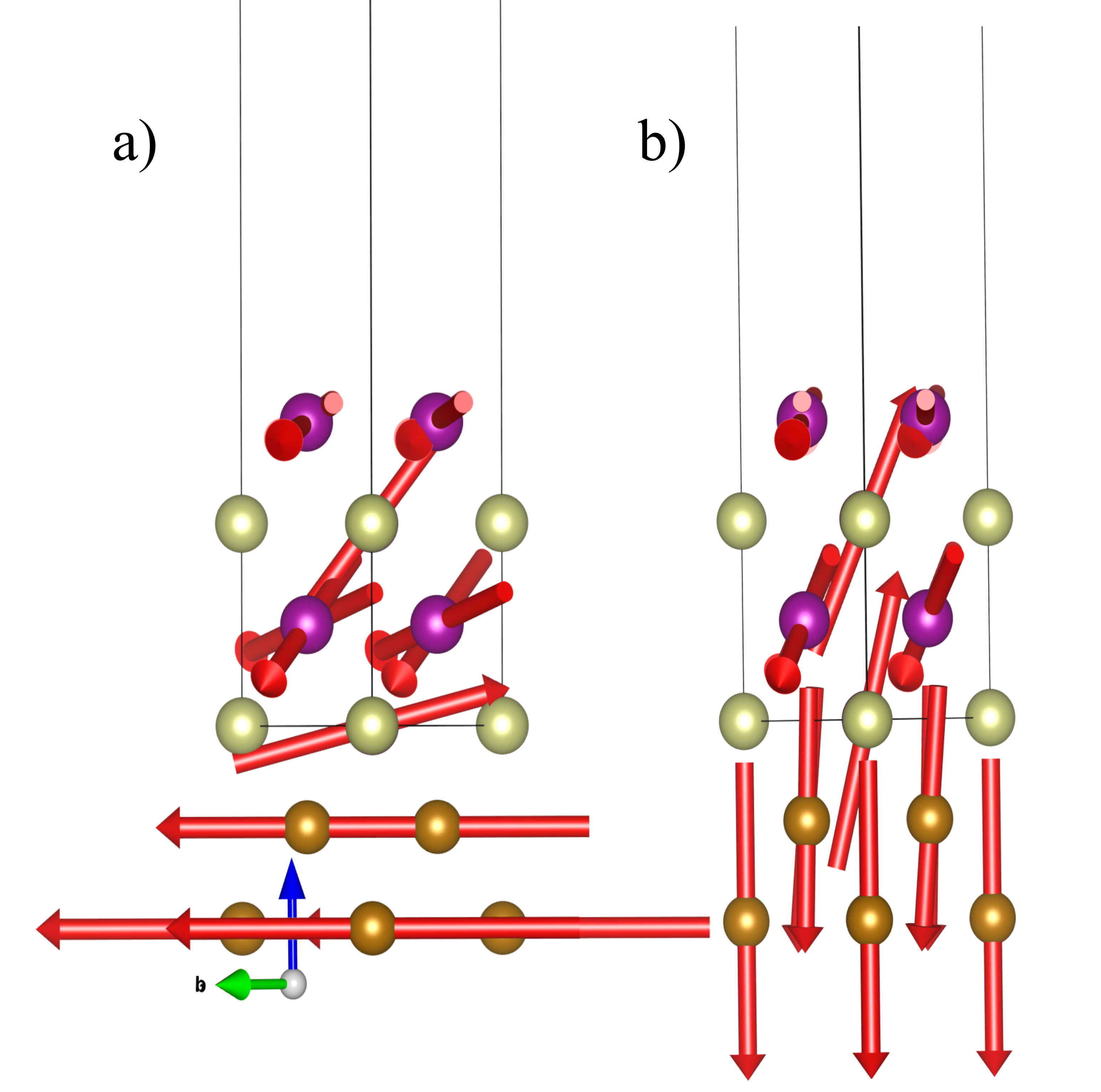}
    \caption{Magnetic structures for the a) Easy and b) Hard magnetic configurations for Fe on the Ir-rich [100] IrMn surface. The easy configuration has a sharper interface between the two separate magnetic structures with the predominant motion being for the spin on the single Mn at the interface. The deviations are relatively small in the hard-axis case, leading to a reduction in the MAE.}
    \label{fig:100_Fe_spins_Ir}
\end{figure}

The [100] structure created in the previous section contains 2 distinguishable interfaces -- one Ir rich, and one Mn rich -- at each face. Accordingly, the Fe layer was independently applied to each surface (such that one surface was always interacting with vacuum) These structures are presented in figure \ref{fig:100_Fe_spins} for the Mn-rich surface, and figure  \ref{fig:100_Fe_spins_Ir} for the Ir-rich surface. The key change here is that the quasi-linear Mn-rich surface (see figure \ref{fig:100_vac}) was restored to a more bulk-like configuration with significant canting into the layer of the IrMn. This confirms that the symmetry-breaking of removing atoms was the driving force behind the rearrangement, whereas even with a \emph{different species of atom} present, the bulk-like magnetic ordering continues provided the underlying symmetry of the crystal is unperturbed.

Additionally, we note that the difference between the easy and hard magnetic configurations on the bulk-like magnetic ordering of either layer are such that the easy configuration has the minimally perturbed magnetic ordering, whereas the magnetic order of either or both layers being significantly distorted from bulk leads to an increased energy (i.e. the hard configuration). From this we may draw a more general principle: when given a choice between two orientations, that which has a surface state that most closely preserves the bulk-like magnetic ordering will be the easier direction.

\FloatBarrier
\subsection{[111] Interface}

\begin{figure}
    \centering
    \includegraphics[width=0.5\linewidth]{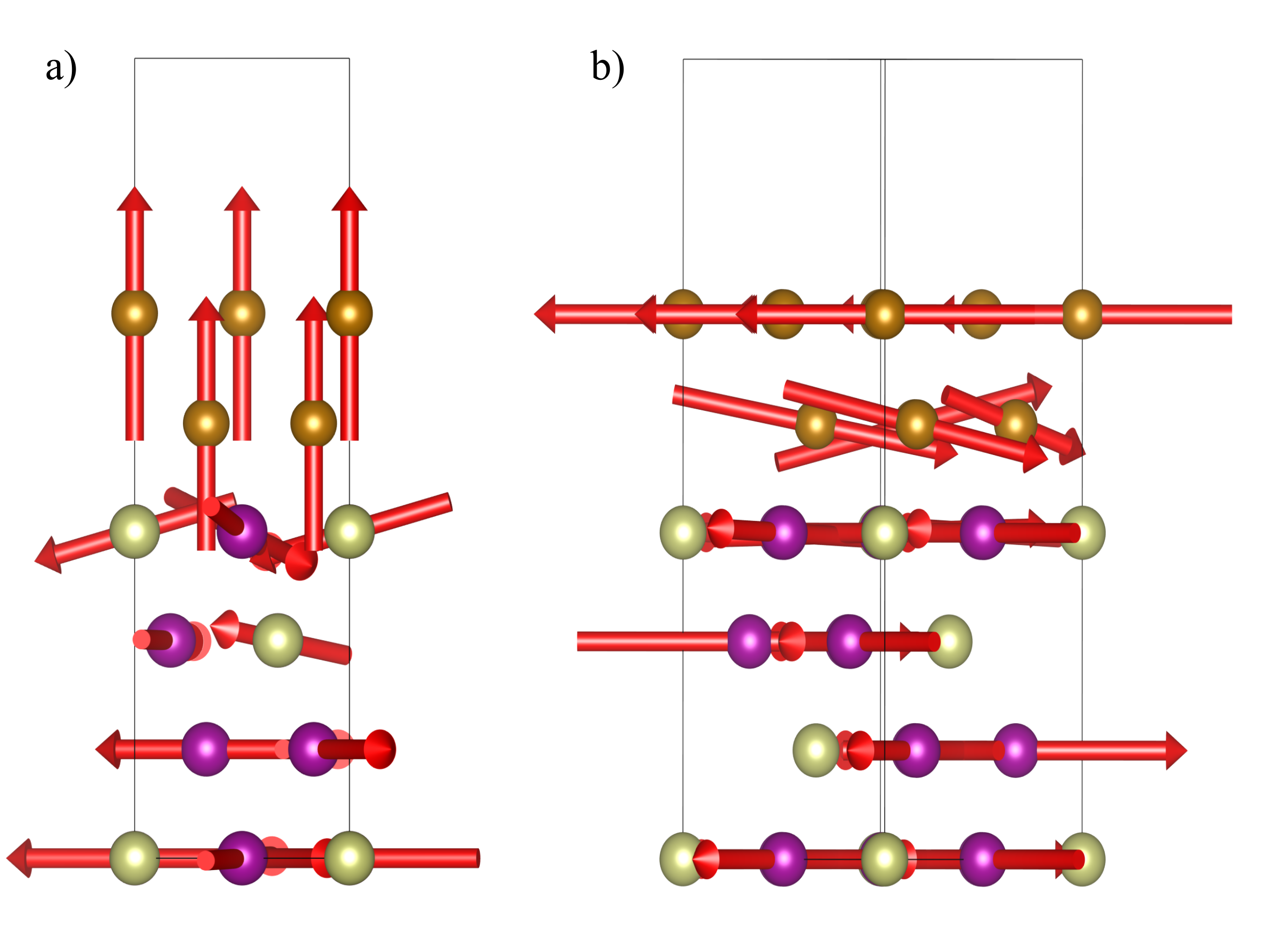}
    \caption{Magnetic structures for the a) Easy and b) Hard magnetic configurations for Fe on a [111] IrMn surface. Note the easy configuration involves a distortion away from the bulk magnetic ordering of IrMn towards the locally easy axes. In panel a, the apparent spins on the Ir atoms are actually on Mn atoms obscured behind them.}
    \label{fig:111_Fe_spins}
\end{figure}

When Fe is added to the [111] surface, there is a distortion of the spin layers in the interface layer only, to form a tetrahedral structure for the magnetic easy axis of the Fe layer (perpendicular to the interface -see panel a of figure \ref{fig:111_Fe_spins}). This leads to a local net moment in the IrMn layer, which is antiparallel but much smaller than the net moment of the Fe layer. When the magnetic moments are in-plane, the bulk-like magnetic ordering is preserved for the IrMn, and the Fe layer directly on top of the IrMn shows the biggest disruption. This indicates that the perpendicular magnetic anisotropy demonstrated by this system is a direct consequence of the surface state formed when adding the Fe causing more stabilisation than leaving the magnetic order of the IrMn layers unperturbed, as we found previously. We may now add an extra \emph{addendum} to our principle from the previous section: \emph{distorting the bulk-like order in such a way that reduces the frustration and makes spin moments orient more closely to their locally preferred (non-interacting) direction will lower the energy of the system}

Conversely, the magnetically hard axis is now oriented in-plane (which was the magnetically easy axis for the [100] projection). Additionally, despite an initial ferromagnetic alignment of the two Fe layers, the IrMn has induced an antiferromagnetic ordering within these two layers, leading to an increase in energy compared with the easy axis. It is probable that this is a surface-effect and would decay back to normal ferromagnetic behaviour with increasing depth of Fe added on top of the IrMn layer. It may also indicate that a bilayer of Fe is too small to allow for proper relaxation in this case. Nevertheless, this result -- while possible surprising -- strongly suggests the importance of the explicit study of magnetic interfaces when predicting both the magnitude and orientation of the exchange bias effect. 

\subsection{Summary of Interfaces with Fe}

Since the Fe atoms have been placed on top of the IrMn$_3$ with the same underlying crystal structure (in order to mimic perfect lattice matching), the underlying symmetry of the crystal structure is only weakly perturbed -- from the perspective of the Mn atom, the change is much smaller between bulk and interfacing with Fe of matching symmetry than between bulk and vacuum. Accordingly, something closely resembling the bulk magnetic ordering is present in the case of the interface with Fe, whereas clear surface-state associated magnetic orderings may be seen in the case of the vacuum interfaces. Broadly, this implies that with sufficiently high-quality lattice matching the underlying bulk magnetic order is possible to maintain. We do note, however, that Fe is a transition state metal adjacent to Mn in the periodic table, and as such has very similar chemical properties -- so replacing Mn with Fe as a nearest neighbour is a small perturbation. Were other elements that were more dissimilar, such as O (common in magnetic oxides) or B (which forms part of FeCoB -- a commonly used capping layer), then this preservation of magnetic order may not persist, especially if there is a poor match, such as with an amorphous seed layer. 

It is worth considering the underlying physics driving the choice of easy axis. In both the [100] and [111] cases, the relative energy of the perpendicular and in-plane configurations for the Fe is determined by how ``smoothly'' the transition occurs between the two layers: the smaller the disruption, the lower the total energy. For the [111] case, the partial lifting of the frustration reduces the energy, and allows a nearly seamless transition between two quasi-bulk like states. 

Together, these indicate that the properties of the system with respect to exchange bias are likely to be dominated by the interface effects, with longer range interactions rapidly decaying. This implies that beyond simple lattice matching, ``magnetic order matching'' must also be considered when designing experimental heterostructure systems in order to select for perpendicular magnetic anisotropy (if desired), or to minimise or maximise coupling between the spin moments in the different materials. This is particularly important in cases, such as IrMn$_3$, where the magnetic frustration makes knowledge of both the local and global minimal orientations important. 

\section{Exchange Bias of Fe}

Magnetocrystalline anisotropy energies (MAE) were calculated for the rotation of the ferromagnetic spin layer of the Fe layers both in the presence and absence of the IrMn layer. The spins were chosen to be either out-of-plane (perpendicular) or in-plane and aligned with the spin-moment of the spin-structure of that plane in the presence of vacuum (parallel). MAE values are calculated as $E_{EB}=E_{Perpendicular}-E_{Parallel}-\Delta E_{Fe}$. where the $\Delta E_{Fe}$ term corrects for the intrinsic anisotropy of the Fe, which is a value on the order of $\mu$eV. A negative value of $E_{EB}$ indicates that the in-plane orientation is hard and the systems exhibits perpendicular magnetic anisotropy, whereas a positive value indicates that the system prefers to lie in-plane. 

\subsection{Choice of Normalisation}

Normalisation is an important tool to compare results fairly between different materials. When dealing with bulk-like systems, the two most widely used normalisation factors are per unit volume, and per formula unit. These both correctly capture the extrinsic nature of magnetocrystalline anisotropy, enabling a fair point of comparison. Both, however, have their limitations. 

Whilst volume normalisation is easily achieved by using x-ray techniques to determine the size of the sample, it does not account for any possible variation in stoichiometry, or for the fact that surface effects will dominate. In theoretical work, it also is not well-defined in the thin film limit as the thickness of the surface is defined by an exponential decay of electron density into the vacuum region, and therefore the volume is entirely dependent on the users choice of threshold for when that has suitably decayed so as to mark the edge of the surface. As a \emph{reductio ad absurdum}, one could consider that if one chose that threshold to be the strict zero of density, then the volume of the thin film would be infinite. 

Formula unit normalisation is more robust with respect to the thin-film thickness problem, and readily comparable between bulk and thin-film theoretical studies, however composition can be experimentally hard to determine, and therefore this makes comparison with experiment less direct. This normalisation is also awkward when non-stoichiometric systems are used, such as our [100] surface. To circumvent this difficulty, we choose to normalise by the number of Mn atoms present in the IrMn$_x$ system.

In this paper we will present both \emph{area} and \emph{per Mn} normalisations, which avoid the problem of the ill-defined surface and should make comparisons with previously published results sufficiently facile. The ratio of Fe:Mn was constant throughout, so no additional insight is gained from normalising by number of Fe atoms.  

\subsection{Magnetocrystalline Anisotropy Energies}

The anisotropy of the Fe layers was calculated to be on the $\mu$eV scale, and therefore below the lowest significant figure for the magnetocrystalline anisotropy energies (MAE) reported in table \ref{tab:MAE}. It is worth noting that the [111] surface has a negative MAE -- this is due to the calculation of the MAE as $E_{Out-of-Plane}-E_{In-Plane}$, so when the in-plane axis is hard (less negative), the sign switches to be negative, indicating the system exhibits so-called ``perpendicular magnetic anisotropy".  

\begin{table}[ht]
    \centering
    \begin{tabular}{c|c|c|c}
        System         & erg/cm$^2$  & meV/$\text{\AA}^2$ & meV/Mn  \\
    \hline
       $[100]$ Ir-rich & 2.08 &   0.13      &  0.31\\
       $[100]$ Mn-rich & 22.3 &   1.39      &  3.33\\
       $[111]$         & -26.0 &  -1.62      & -3.33\\
    \end{tabular}
    \caption{Magnetic anisotropy values for the three surfaces investigated in this work. Negative values indicate perpendicular magnetic anisotropy (in-plane being a hard plane). Note the [100] surface has a significantly suppressed anisotropy at the Ir-rich surface compared with the Mn-rich layer. Two normalisation schemes are included for convenience, with areal normalisation being given in two separate sets of units (1 meV/$\text{\AA}^2$ = 16.02 erg/cm$^2$).}
    \label{tab:MAE}
\end{table}

It is also notable that the exchange bias effect is much weaker for the Ir-rich [100] surface than either of the other two surfaces considered in this study. This is indicative that the spin-spin coupling between the layers rather than the biasing of the magnetic moments of the Fe by the presence of Ir is critical to the exchange bias effect. This is also an optimistic result for the search to replace Ir with less-scarce alternatives, since it indicates that the high-spin of the Mn is more important than the direct presence of the Ir for achieving a high anisotropy.

Finally, we note that there are two competing mechanism driving exchange bias. Firstly, one can have a ``fixed'' AFM layer, where the interfacial layer of the capping material conforms wholly or partially to an unperturbed bulk-like magnetic structure of the AFM. In this case, the increase in energy is almost entirely internal to the FM, and caused by Heisenberg J-couplings and 2 body anisotropies (i.e. the misalignment of the spins). By changing the orientation of the capping layer, the degree of misalignment may be minimised, defining the magnetically easy direction. This sort of mechanism was apparent in the [100] based interface (figures \ref{fig:100_Fe_spins} and \ref{fig:100_Fe_spins_Ir}).  

Alternatively, we also see an alternative mechanism in the [111]-based case. An appropriate magnetic alignment between the two layers can reduce the level of frustration present in the surface layer. This reduces the energy of the system. Conversely, an unsympathetic alignment recovers the previous mechanism, which tends to give the minimally destabilising rather than an actively stabilising effect. We note that within our ultrathin layers, there is only one layer of magnetic ions within the ferromagnet which are free to rotate. This compresses the distance over which the spins may relax, which could have the effect of increasing the reported anisotropies. conversely, we anticipate that the metallic nature of the materials are likely to provide significant screening, such that the distortions to the magnetic ordering converge back to bulk within only a few layers, such that this compression is not unreasonable. For insulating magnetic materials (not considered in this work) a more thorough investigation of MAE vs capping layer thickness is likely to be critical.

It is also worth noting that the maximum amount of achievable exchange bias is not dependent purely on the intrinsic anisotropy of the antiferromagnet. The kinetics of reversal (MAE is the activation energy for this process), are dominated by the minimum energy pathway due to the exponential factor in the N\'eel-Arrhenius equation. This pathway \emph{can} leave the antiferromagnet unaffected if disrupting the internal magnetic order of the ferromagnetic capping layer has a lower anisotropy. Since the ferromagnetic layers in real devices have properties which tend towards bulk-like, this suggests that above a certain critical anisotropy within the antiferromagnet, the reversal mechanism is dominated by the magnetic properties of the capping layer. Accordingly, to investigate the exchange bias effect, theoretical studies ought to consider interfacial effects explicitly in order to appropriately capture the physics of the real system.

\section{Conclusion}

Magnetic structures of IrMn$_3$ terminated systems were studied using a novel method based on spin initialisation determined by the local point group of the atom in question. These predicted a different, lower-energy, spin structure for the surface layer when interfaced with vacuum due to the especially high degree of symmetry breaking in this case. Confinement of the disruption of the bulk-like magnetic ordering to the surface layer may be driven by the metallicity of the system. The precise magnetic structure depended on the choice of termination. When capped with Fe, the bulk-like spin structure was much more noticeable, even at the interface layer. This indicates that the triangular magnetic ordering of the bulk phase is both stable down to very thin (1-2 unit cells thick) films, and that determining the magnetic structure at the interface is sensitive to what the interface is between; with larger deviations from bulk-like order leading to larger changes in the magnetic ordering. Finally, the magnetocrystalline anisotropy energies per unit area were calculated and determined to be 1.39, 0.13, and -1.62 meV/$\text{\AA}^2$ for the [100] Mn-rich, [100] Ir-rich and [111] surfaces.  

\section{Acknowledgements}

The authors acknowledge EPSRC grant EP/V047779/1 for financial support. We are grateful for computational support from the UK national high performance computing service, ARCHER2\cite{Archer2}, for which access was obtained via the UKCP consortium and funded by EPSRC grant ref EP/X035891/1

\section{Supplementary Data}
Input files for simulations reported in this paper is available for download from the research data repository of the University of York at DOI: 10.15124/d9c9ef10-036e-4bb5-ab6c-ba8b00207668

\bibliographystyle{unsrt}
\bibliography{bib}

\begin{thebibliography}{10}

\bibitem{Misiorny2013}
Maciej Misiorny, Michael Hell, and Maarten~R. Wegewijs.
\newblock Spintronic magnetic anisotropy.
\newblock {\em Nature Physics}, 9(12):801–805, October 2013.

\bibitem{Samanta2024}
Kartik Samanta, Yuan-Yuan Jiang, Tula~R. Paudel, Ding-Fu Shao, and Evgeny~Y. Tsymbal.
\newblock Tunneling magnetoresistance in magnetic tunnel junctions with a single ferromagnetic electrode.
\newblock {\em Physical Review B}, 109(17), May 2024.

\bibitem{Giri2011}
S~Giri, M~Patra, and S~Majumdar.
\newblock Exchange bias effect in alloys and compounds.
\newblock {\em Journal of Physics: Condensed Matter}, 23(7):073201, February 2011.

\bibitem{Frost2024}
William Frost, Fatimah Alsaud, Robert~A. Lawrence, Matt Probert, and Gonzalo~Vallejo Fernandez.
\newblock Towards mnn as a replacement for irmn.
\newblock {\em Scientific Reports}, 14(1), September 2024.

\bibitem{Vesborg2012}
Peter C.~K. Vesborg and Thomas~F. Jaramillo.
\newblock Addressing the terawatt challenge: scalability in the supply of chemical elements for renewable energy.
\newblock {\em RSC Advances}, 2(21):7933, 2012.

\bibitem{Kang2021}
Jaimin Kang, Jeongchun Ryu, Jong-Guk Choi, Taekhyeon Lee, Jaehyeon Park, Soogil Lee, Hanhwi Jang, Yeon~Sik Jung, Kab-Jin Kim, and Byong-Guk Park.
\newblock Current-induced manipulation of exchange bias in irmn/nife bilayer structures.
\newblock {\em Nature Communications}, 12(1), November 2021.

\bibitem{Carter2024}
Alexandra Carter, Samridh Jaiswal, Paolo Campiglio, and Gonzalo Vallejo-Fernandez.
\newblock Influence of composition on the magnetisation reversal of irmn/cofe exchange bias systems.
\newblock {\em Journal of Magnetism and Magnetic Materials}, 598:172035, May 2024.

\bibitem{Bufaial2024}
L.~Bufai\c{c}al and E.M. Bittar.
\newblock Essential aspects of the spontaneous exchange bias effect.
\newblock {\em Journal of Magnetism and Magnetic Materials}, 599:172109, June 2024.

\bibitem{Aley2010}
N.~P. Aley, M.~Bowes, R.~Kr\"{o}ger, and K.~O’Grady.
\newblock Texture and magnetic properties of exchange bias systems.
\newblock {\em Journal of Applied Physics}, 107(9), May 2010.

\bibitem{Peng2020}
Shouzhong Peng, Daoqian Zhu, Weixiang Li, Hao Wu, Alexander~J. Grutter, Dustin~A. Gilbert, Jiaqi Lu, Danrong Xiong, Wenlong Cai, Padraic Shafer, Kang~L. Wang, and Weisheng Zhao.
\newblock Exchange bias switching in an antiferromagnet/ferromagnet bilayer driven by spin–orbit torque.
\newblock {\em Nature Electronics}, 3(12):757–764, November 2020.

\bibitem{Fan2022}
R.~Fan, R.O.M. Aboljadayel, Alexey Dobrynin, Peter Bencok, R.C.C. Ward, and Paul Steadman.
\newblock Dependence of exchange bias on structure of antiferromagnet in fe/irmn$_3$.
\newblock {\em Journal of Magnetism and Magnetic Materials}, 546:168678, March 2022.

\bibitem{Bhattarai2025}
Romakanta Bhattarai, Peter Minch, and Trevor~David Rhone.
\newblock High-throughput screening of altermagnetic materials.
\newblock {\em Physical Review Materials}, 9(6), June 2025.

\bibitem{Ceulemans1984}
A.~Ceulemans, D.~Beyens, and L.~G. Vanquickenborne.
\newblock Symmetry aspects of jahn-teller activity: structure and reactivity.
\newblock {\em Journal of the American Chemical Society}, 106(20):5824–5837, October 1984.

\bibitem{Ascher1977}
E~Ascher.
\newblock Permutation representations, epikernels and phase transitions.
\newblock {\em Journal of Physics C: Solid State Physics}, 10(9):1365–1377, May 1977.

\bibitem{Prodan2005}
E.~Prodan and W.~Kohn.
\newblock Nearsightedness of electronic matter.
\newblock {\em Proceedings of the National Academy of Sciences}, 102(33):11635–11638, August 2005.

\bibitem{Clark2005}
Stewart~J. Clark, Matthew~D. Segall, Chris~J. Pickard, Phil~J. Hasnip, Matt I.~J. Probert, Keith Refson, and Mike~C. Payne.
\newblock First principles methods using castep.
\newblock {\em Zeitschrift f\"{u}r Kristallographie - Crystalline Materials}, 220(5–6):567–570, May 2005.

\bibitem{Perdew1992}
John~P. Perdew and Yue Wang.
\newblock Accurate and simple analytic representation of the electron-gas correlation energy.
\newblock {\em Physical Review B}, 45(23):13244–13249, June 1992.

\bibitem{Perdew2018}
John~P. Perdew and Yue Wang.
\newblock Erratum: Accurate and simple analytic representation of the electron-gas correlation energy [phys. rev. b 45, 13244 (1992)].
\newblock {\em Physical Review B}, 98(7), August 2018.

\bibitem{Pulay1980}
Péter Pulay.
\newblock Convergence acceleration of iterative sequences. the case of scf iteration.
\newblock {\em Chemical Physics Letters}, 73(2):393–398, July 1980.

\bibitem{OGrady2020}
K.~O’Grady, J.~Sinclair, K.~Elphick, R.~Carpenter, G.~Vallejo-Fernandez, M.~I.~J. Probert, and A.~Hirohata.
\newblock Anisotropy in antiferromagnets.
\newblock {\em Journal of Applied Physics}, 128(4), July 2020.

\bibitem{Cuadrado2018}
R~Cuadrado, M~Pruneda, A~García, and P~Ordejón.
\newblock Implementation of non-collinear spin-constrained dft calculations in siesta with a fully relativistic hamiltonian.
\newblock {\em Journal of Physics: Materials}, 1(1):015010, November 2018.

\bibitem{Szunyogh2009}
L.~Szunyogh, B.~Lazarovits, L.~Udvardi, J.~Jackson, and U.~Nowak.
\newblock Giant magnetic anisotropy of the bulk antiferromagnets $irmn and irmn_3$from first principles.
\newblock {\em Physical Review B}, 79(2), January 2009.

\bibitem{Jenkins2021}
Sarah Jenkins, Roy.~W. Chantrell, and Richard F.~L. Evans.
\newblock Atomistic simulations of the magnetic properties of $ir_xmn_{1-x}$ alloys.
\newblock {\em Physical Review Materials}, 5(3), March 2021.

\bibitem{OGrady2010}
K.~O’Grady, L.E. Fernandez-Outon, and G.~Vallejo-Fernandez.
\newblock A new paradigm for exchange bias in polycrystalline thin films.
\newblock {\em Journal of Magnetism and Magnetic Materials}, 322(8):883–899, April 2010.

\bibitem{Archer2}
George Beckett, Josephine Beech-Brandt, Kieran Leach, Z\"{o}e Payne, Alan Simpson, Lorna Smith, Andy Turner, and Anne Whiting.
\newblock Archer2 service description, 2024.

\end{thebibliography}
\end{document}